\documentclass[11pt,twoside,notitlepage,a4paper,twocolumn]{article}
\usepackage{amssymb} % amsfonts 
\author{K.-H. Rehren\footnote{address after Nov.\ 1997: 
Institut f\"ur Theoretische Physik, Universit\"at G\"ottingen (Germany)} \\
II.\ Institut f\"ur Theoretische Physik, \\ Universit\"at Hamburg 
(Germany)}
\title{\vskip-14mm Spin-Statistics and CPT for Solitons}     
 \def\eps{\varepsilon} \def\a{\alpha} \def\b{\beta} 
\def\g{\gamma} \def\Kappa{\mbox{\boldmath $\kappa$}} 
\def\Mu{\mbox{\boldmath $\mu$}}
\def\inv{{^{-1}}}  \def\comp{{\scriptstyle\circ}} \def\Y{{\mathbf Y}}
\def\frac#1#2{{#1\over#2}}  \def\pair#1#2{{\langle#1,#2\rangle}}
\def\Ei{\hbox{\rm 1}\mskip-4.4mu\hbox{\rm l}}
\def\Ad{\hbox{\rm Ad}\,} \def\id{\hbox{\rm id}} 
\def\RR{\mathbb{R}} \def\ZZ{\mathbb{Z}} 
\def\qed{\hfill$\square$} \def\bea{\begin{array}} \def\eea{\end{array}} 
\begin{document}
\pagestyle{empty} 

%\begin{center} \vskip10cm \Large Spin-Statistics and
%  CPT for solitons \\[3cm] \large Karl-Henning Rehren
%\\[3cm] \rm November 1997 \end{center} \newpage 

%${}$\newpage ${}$\newpage ${}$\newpage ${}$\newpage 
\pagestyle{headings} \setcounter{page}{1}
\parskip1mm plus.4mm minus.2mm \parindent0mm 
\textheight231mm \textwidth170mm \topmargin0mm 
\evensidemargin-5mm %\oddsidemargin0mm 
\renewcommand{\today}{}
\maketitle
{\bf Abstract:} {\small
It is analysed under which conditions the statistics of soliton sectors
of massive two-dimensional field theories can be properly defined. 
A soliton field algebra is defined as a crossed product with the group 
of soliton sectors. In this algebra, the non-local commutation relations are 
determined and Weak Locality, Spin-Statistics and CPT theorems are proven. 
These theorems depart from their usual appearance due to the broken symmetry 
connecting the inequivalent vacua. An interpretation of these results in terms 
of modular theory is given. For the neutral subalgebra of the soliton algebra, 
the theorems hold in the familiar form, and Twisted Locality is derived.}

\section{Introduction} 

Solitons naturally appear in two-dimensional quantum field theory as
superselection sectors of the observable algebra whenever there are
several inequivalent vacuum sectors. The latter may arise due to the
spontaneous breakdown of some (discrete) symmetry. In this generality, the
notion of solitons is not related to integrability, and was first discussed
by Fr\"ohlich \cite{JF} and later, without reference to a broken symmetry, 
by Fredenhagen \cite{KF}.
It is true that in all known models the degeneracy of the vacua is due to
a broken abelian symmetry, and therefore we shall concentrate on this case.

It is the purpose of this letter to initiate a study of the fundamental
theorems of quantum field theory such as Spin-Statistics and CPT for solitonic 
fields. In fact, we shall see that under suitable conservative assumptions, 
these theorems can be maintained but in a modified form. E.g., neither the spin
nor the commutation relations of solitonic fields are determined by the 
statistics, as was emphasizd already by Schroer and Swieca \cite{SS},
but depend also on other quantum numbers related to the broken symmetry.

It is of prime interest to control the departure from the naive form of the
theorems in the most general case. But this is not the purpose of the 
present letter, and would inflate its dimensions. Instead, we want to
emphasize the automatic appearance of modifications even in the most simple 
cases. This attitude should justify several simplifying conditions and 
assumptions made in the course of the arguments. Most of these are indeed 
easily proven in the case of $\ZZ$- or $\ZZ_N$-symmetries, and presumably
with some effort also for more general symmetry groups, as will be pointed 
out along the way. We shall also omit many technical details and rather
concentrate on the necessary {\sl modifications} of standard 
treatments of Spin-Statistics and CPT \cite{BW,BSM,FRS2}.

Our first topic will be the question whether the intrinsic notion of 
statistics due to Doplicher, Haag, and Roberts (DHR) \cite{DHR1,DHR2} extends 
to the case of soliton sectors at all. Having settled this issue, we proceed
to the construction of the soliton algebra which extends the observable 
algebra and contains charged fields which interpolate between all
soliton sectors of the observables. This canonical construction just 
copies the construction for simple DHR sectors \cite{DHR1}. But in 
contrast to the DHR case, the soliton algebra not only transforms under an
unbroken gauge symmetry, but also inherits the broken symmetry of the 
observables, thereby giving rise to two sets of ``quantum numbers'', one
being superselected and the other not. 
  
Our main results are the derivation of the space-like commutation relations 
for the soliton algebra, of the Spin-Statistics and CPT theorems and the 
relation of the latter to a modular structure reminiscent of the 
Bisognano-Wichmann theorem \cite{BW}. The formal difference from the standard 
theorems is an ubiquitous dependence on the broken quantum numbers. Especially 
for the commutation relations, this must have been anticipated since the 
soliton fields asymptotically implement the symmetries of the observables. 
But the deep algebraic difference lies in the fact, that the solitonic modular 
operator is not a Tomita operator (whose modular conjugation maps an algebra 
into its commutant), and therefore the soliton commutation relations, unlike 
anyonic or plektonic ones \cite{FRS2}, cannot be interpreted as a twisted 
commutativity.

Twisted commutativity turns out to be restored upon the passage to the
algebra of ``neutral'' solitons which are the fixpoints under the broken 
symmetry. But, as the discussion in the last section will explain, this
algebra is indeed the field algebra associated with the neutral observables 
via extension by the {\sl anyonic} DHR sectors of the latter. 

A prime motivation for the present study is a recent preprint of Niedermaier
\cite{MN} who endeavours to derive the Cyclic Form Factor (CFF) Equation, which
is usually considered as an axiom in the form factor approach to integrable
models, from an asserted KMS property of the vacuum states on the soliton 
algebra with respect to the Lorentz boosts, in conjunction with the 
Haag-Ruelle scattering theory but without recourse to integrability. 
For this purpose, Niedermaier had to {\sl postulate} the underlying modular 
structure, and indeed in the first version of his paper ignored the role of 
the broken symmetry quantum numbers as described above. 
By our present results, Niedermaier's line of
reasoning can be justified either for the full soliton algebra with some
modifications, or for the neutral solitons without modifications. The
modifications all accumulate in the phase factor $\eta$ 
entering the CFF equation. This factor was left more or less unspecified,
anyway, but was asserted to arise from the commutation relations alone only 
while our present analysis shows that it has additional contributions from the
modified modular operator.

\section{Statistics of solitons}

In massive theories\footnote{to be precise: in theories satisfying the 
split property for wedges (SPW). SPW is expected - but was 
proven only in 
simple examples - to be a manifestation of a mass gap.} with Haag duality, 
it was shown by M\"uger \cite{MM2} that nontrivial DHR sectors do not exist. 
Hence all positive energy representations of the observable algebra $A$
are either vacuum representations or, provided there are several vacua, 
soliton representations asymptotically connecting two different vacua. When 
the vacuum degeneracy is due to a spontaneously broken symmetry \cite{JF}, 
then any two vacuum representations are related as $\pi_\b = \pi_\a \comp g$ 
where $g\in G$ is a global discrete symmetry transformation commuting with 
the Poincar\'e group.

Then \cite{JF,DS} for every pair $\a,\b$ there is a soliton representation 
which on the right complement of some double-cone concides with $\pi_\a$, 
and on its left complement coincides with $\pi_\b$.

All soliton representations can be modelled on the same Hilbert space of some 
reference vacuum sector $H_\a$ with the help of soliton automorphisms 
$g \leftarrow \rho \rightarrow h$ of $A$ evaluated in the vacuum 
representation $\pi_\a$.\footnote{The notation means that on the left and 
right space-like complements of some bounded interpolation region, $\rho$ 
acts like $g$ and like $h$, respectively. The interpolation regions may be 
assumed to be double-cones, 
that is, open intersections of a forward and a backward light-cone.} 
Among them are, of course, also the other vacua since 
$g \leftarrow g \rightarrow g$ is a special case of a soliton automorphism. 
The automorphisms $\rho$ can be freely composed.

The DHR approach to superselection sectors and statistics
\cite{DHR1,DHR2}\footnote{We assume some familiarity with this theory and 
shall refrain from repeating the lengthy but standard arguments throughout 
this article. A rather compact introduction can be found in \cite{FRS1}.} 
addresses the question whether $\rho_1\rho_2$ is unitarily equivalent with 
$\rho_2\rho_1$. Looking at the asymptotic behaviour, it is clear that the 
group $G$ must be necessarily abelian.

The issue is now whether the product of two soliton automorphisms is 
commutative whenever their interpolation regions are at space-like distance. 
Since for every $\rho$, every local observable $a$ and every $k \in G$ 
one can choose an appropriately translated soliton automorphism 
$\sigma$ which equals $k$ on $a$ and on 
$\rho(a)$, requiring $\rho$ to commute with $\sigma$ at space-like distance
implies that necessarily $\rho k = k \rho$ on all $a \in A$.

We therefore impose the condition on the soliton automorphisms (which also 
implies that $G$ is abelian) \\[1mm]
(A) \hfil $\rho k = k \rho\qquad (k \in G)$. \hfil  

It is not obvious that, given $G$ abelian, every soliton
sector has a representative satisfying (A). A rather general proof is 
announced by M\"uger \cite{MM3}.

The next issue is transportability. For $\rho$ in a covariant sector, there 
is a unitary cocycle (with respect to the 
Poincar\'e group) of intertwiners $\g_\rho(x,\lambda):\rho\rightarrow\hat\rho=
\a_{x,\lambda}\rho\a_{x,\lambda}\inv$ (charge transporters). By Haag duality,
these are localized operators, but for the application of the DHR theory
a little more is required, namely that soliton automorphisms act trivially 
on charge transporters at space-like distance. Hence we need that charge 
transporters are invariant under $G$ (``neutral''). Indeed, we have:

{\bf Lemma 1:} \sl If the symmetry group $G$ commutes with the 
Poincar\'e group $P$ on the observables, and if $\rho$ commutes with $G$, 
then the charge transporters $\g_\rho(x,\lambda)$ for the translations 
($\lambda = 1$) are neutral. If $G$ is finite, then also the 
charge transporters for the boosts are neutral. \rm

{\sl Proof:} $\hat\rho$ again commutes with $G$, and hence along with
$\g_\rho(x,\lambda)$ also $k(\g_\rho(x,\lambda))$ is another cocycle with the 
same intertwining property. The two can therefore only differ by an 
$x,\lambda$- and $k$-dependent phase which must be a representation of both 
the Poincar\'e group and $G$, or put differently, a representation of $P$ with
values in $\hat G$. If $G$ is finite, then $\hat G$ is discrete, and 
by continuity, the phase must be trivial. If $G$ is not finite, then
for every $k\in G$, the phase is a one-dimensional representation of $P$. 
But all such representations are trivial on the translations. \qed

For $G$ not finite, possibly constructive methods \cite{DS} to obtain the 
cocycle give neutrality for the boosts also. In this section we shall need 
only the translations, but for later purposes, we shall have to impose the 
condition 

(B) \sl The cocycles $\gamma_\rho(\lambda)$ for the boosts are neutral.\rm 

It is now easy to show with the standard DHR arguments \cite{DHR2}, using
neutral charge transporters whenever necessary, that covariant soliton 
automorphisms satisfying condition (A) indeed commute at space-like distance, 
and commute up to unitary equivalence in general.

We want to proceed as usual \cite{DHR2} and define unitary statistics 
operators $\eps(\rho_1,\rho_2):\rho_1\rho_2 \rightarrow \rho_2\rho_1$ 
for any pair of covariant soliton automorphisms by first translating them 
with the help of unitary charge transporters until they are 
at space-like distance, then commuting them and translating back. 
Since this strategy involves the application of soliton automorphisms 
to auxiliary charge transporters at space-like distance, the independence on 
the choice of the latter is assured only if they are neutral.
The statistics operator would be sensitive to any perturbation by a 
charge transporter which is not neutral \cite{SS}.

It is now just a repetition of the DHR arguments \cite{DHR2,FRS1} to show that 
the statistics operators are well-defined if one admits only neutral 
charge transporters (which exist by Lemma 1), and that they depend 
at most on the orientation of the auxiliary regions at space-like distance 
chosen in the process. By convention, the possibility with $\rho_1$ 
transported to the right of $\rho_2$ defines $\eps(\rho_1,\rho_2)$, and
consequently the opposite choice yields $\eps(\rho_2,\rho_1)^*$.
One easily derives

{\bf Proposition 1:} \sl The statistics operators for soliton automorphisms
satisfying {\rm (A)} defined with neutral charge transporters are
themselves neutral. They satisfy naturality, e.g., 
$$ u_2\eps(\rho_1,\rho_2) = \eps(\rho_1,\hat\rho_2)\rho_1(u_2) $$
with neutral intertwiners $u_2:\rho_2\rightarrow\hat\rho_2$
and multiplicativity, e.g.,
$$ \eps(\rho_1,\rho_2\rho_3) = \rho_2(\eps(\rho_1,\rho_3))\eps(\rho_1,\rho_2) 
$$
in both entries. They satisfy the braid group identities. The
statistics of $\rho = k \in G$ is trivial:
$$ \eps(k,\rho_2) = \eps(\rho_1,k) = \Ei .$$
The monodromy operator is given by
$$ \eps(\rho_1,\rho_2)\eps(\rho_2,\rho_1) = \frac{\kappa_{\rho_1\rho_2}}
{\kappa_{\rho_1}\kappa_{\rho_2}} \cdot \Ei. $$ \rm

Here, since we are dealing with {\sl auto}morphisms, the statistics operator 
$\eps(\rho,\rho)$ is just a complex phase called the statistics phase 
$\kappa_\rho$. It is invariant under perturbations of $\rho$
by any neutral unitary. For the braid group identity, cf.\ \cite{FRS1}.

{\bf Corollary:} $\kappa_\rho = \kappa_{\rho\inv} = \kappa_{\rho\comp k}$ for
$k \in G$.

The present results do not exclude the possibility that the 
group of soliton automorphisms generated by some $\rho$ contains some 
$\hat\rho$ which is unitarily equivalent to $\rho$ but not related by a 
neutral unitary. E.g., imagine $\rho^2 = \Ad u$ where $u$ is not 
neutral. (Since $\rho^2$ commutes with $G$, it is only
guaranteed that $g(u)u^*$ is a scalar.) Therefore, the statistics within
any group of soliton automorphisms containing $\rho$ might not be an invariant 
of the sector ($\rho^3 = \Ad u\comp \rho$ will have different statistics 
from $\rho\,$!).

We exclude this possibility by picking one representative
$g \leftarrow \rho_{g,h} \rightarrow h$ for every pair of vacuum
sectors, with the property

(C) \sl The assignment $(g,h) \mapsto \rho_{g,h}$ is
a group homomorphism $G \times G \to Aut(A)$, and $\rho_{g,g}=g \in G$. \rm
 
This property entails condition (A) if $G$ is abelian. It is easily
fulfilled for $G = \ZZ$ by choosing 
$\rho_{e,g}$ to be powers of some generating element commuting with
$G$, and putting $\rho_{h,g}=g\rho_{e,hg\inv}$.
M\"uger informed us that he can in fact {\sl construct} such a homomorphism
also in the general case \cite{MM3}. 

Within the group of soliton automorphisms $\{\rho_{g,h}\}$, the
statistics has all the properties of the statistics of 
low-dimensional simple DHR sectors \cite{DHR2,FRS1}.
As a consequence of property (C), all statistics operators 
between $\rho_{g,h}$ are just numerical phases since the $\rho_{g,h}$ 
commute with each other, and as a consequence of Prop.\ 1 they are 
multiplicative in both entries as well as invariant under shifts by
$k \in G$. We summarize some further properties for later reference.

{\bf Lemma 2:} (i) \sl Let $g_i \leftarrow \rho_i \rightarrow h_i$, $i=1,2$. 
Then
$$ e(k_1,k_2) := \eps(\rho_{e,k_1},\rho_{e,k_2}) = \eps(\rho_1,\rho_2) $$
depends only on $k_i = h_ig_i\inv$. This function on $G \times G$ is a 
two-cocycle, that is
$$ e(k_1,k_2)e(k_1k_2,k_3) = e(k_1,k_2k_3)e(k_2,k_3) .$$

{\rm (ii)} If $e(k_1,k_2)$ is a coboundary (which is automatic if $G$ has 
trivial second cohomology), that is
$$ e(k_1,k_2) = \frac{\mu(k_1k_2)}{\mu(k_1)\mu(k_2)}, $$
then one can choose the complex phases $\mu(k)$ to satisfy 
$$\mu(e) = 1 \quad \hbox{and}\quad \mu(k) = \mu(k\inv) $$
$$\mu(k)^2 = e(k,k) \equiv \kappa_{\rho_{g,h}} \quad \hbox{if} 
\quad k=hg\inv. $$

Proof: \rm The first statement in (i) is $G$-invariance, and the second 
statement follows from multiplicativity of the statistics operators (Prop.\ 1).
The coboundary property in (ii) already 
implies $\mu(e) = 1$ for $k_2=e$, and $\mu(k)\mu(k\inv) = 
e(k,k)$ for $k_1k_2 = e$. The possibility to choose $\mu(k) = 
\mu(k\inv)$ is seen as follows. Let $G$ be generated by 
commuting elements $a_i$.  
We choose a square root $\mu_i$ of $e(a_i,a_i)$ for every generator and put
$$ \mu(k) := {\prod}_i \mu_i^{\;p_i^2} \; 
{\prod}_{i<j} e(a_i,a_j)^{p_ip_j} $$
if $k = \prod_i a_i^{p_i}$ is a representation in terms of the generators, 
with the qualification that $p_i$ has to be chosen even if $a_i$ is of
odd order. This ensures the definition to be independent of the choice of 
$p_i$ since $\mu_i^{2N} = 1 = e(a_i,a_j)^N$ whenever $a_i^N = 1$. 
Furthermore, direct computation shows that the coboundary of $\mu$ thus
defined yields $e$, using $e(a_1,a_2) = e(a_2,a_1)$ (being a coboundary) 
and multiplicativity of $e$ in both entries (Prop.\ 1)\footnote{This
argument actually shows that a multiplicative two-cocycle is a coboundary if 
and only if it is symmetric.}. \qed 

\section{The soliton algebra}

In this and the next sections, the conditions (A)--(C) shall 
be understood.

Property (C) is of structural importance since it permits to define 
the soliton algebra as the crossed 
product of the observables with the group of soliton sectors $G \times G$ 
acting through the automorphisms $\rho_{g,h}$ \cite{DHR1}, or equivalently as
a global section through the field bundle \cite{DHR2}, as follows. 

All superselection sectors of the observables are realized by $\pi_\a\comp\rho$
on the representation space of a reference vacuum sector $\pi_\a$, and 
$\rho = \rho_{g,h}$ exhaust the irreducible equivalence classes. Calling the
Hilbert space equipped with these representations $H_{g,h}$, we define the 
soliton algebra $B$ acting irreducibly on the direct sum 
$H = \bigoplus_{g,h} H_{g,h}$. 

It is generated by its elements $F = (\rho_{g,h},b)$ ($b \in A$) acting on 
each subspace by
$$ (\rho,b) (\pi,\Phi) = (\pi\rho,\pi(b)\Phi) .$$
It contains the observables $a \equiv (\id,a)$ as well as unitary field
operators which implement the automorphisms $\rho_{g,h}$ of the observables.

The algebraic structure of $B$ is determined by the multiplication law
$$ (\rho_1,b_1)(\rho_2,b_2) = (\rho_2\rho_1,\rho_2(b_1)b_2) $$
which shows that $B$ is a crossed product of $A$ with the group of
soliton sectors. The adjoint in $B$ is
$$ (\rho,b)^* = (\rho\inv,\rho\inv(b^*)) $$
and makes $B$ a C* algebra. 

Let the representatives $\rho$ be localized in some region $O$. Then
$F=(\rho,b)$ is localized in $\lambda O+x$ if $\g_\rho(x,\lambda)b 
\in A(\lambda O+x)$.
This definition of localization guarantees solitonic commutation relations 
with the observables at space-like distance, and is preserved by the * 
operation.

The soliton algebra $B$ is Poincar\'e covariant with
$$\a_{x,\lambda}(\rho,b) = (\rho,\g_\rho(x,\lambda)^*\a_{x,\lambda}(b)) , $$
extending the Poincar\'e transformations of $A \subset B$.

The soliton algebra possesses two (commuting) internal symmetries: first 
there is an unbroken dual gauge symmetry
$$F \stackrel\eta\mapsto \eta(g,h) \cdot F$$ 
if $F$ belongs to the sector $(g,h)$, where $\eta \in \hat G \times \hat G$ 
is a character of $G \times G$. Second, the original broken symmetry extends 
to $B$ by the action of $k \in G$
$$ (\rho,b) \stackrel k\mapsto (\rho,k(b)). $$
Both symmetries preserve localization and commute with *. Actually, the 
$G$-symmetry is now {\sl inner}: it is implemented by the unitaries $Y_k^*$ 
where $Y_k = (k,\Ei) \in \bigcap_O B(O)$ connect the vacua by $Y_k\Omega_\a = 
\Omega_{\a\comp k}$.

We shall call $\chi \in \hat G$ the ``character'' of a field operator 
whenever the latter transforms like
$$ k(F) = \pair \chi k \cdot F .$$

The dual symmetry commutes with the Poincar\'e group. The
$G$-symmetry commutes with the translations, and also with the boosts
only provided the cocycle is neutral (cf.\ condition (B) and Lemma 1). 
Otherwise $\a_\lambda\comp k$ and $k \comp \a_\lambda$ would differ by a 
continuous dual gauge transformation.

Now, repeating the reasoning as in the case of DHR sectors \cite{FRS1}, 
one finds the commutation relations at space-like distance:
 
{\bf Proposition 2:} (Commutation Relations) \sl Let $F_i = (\rho_i,b_i)$ be 
localized in $O_i$ at space-like distance, with $g_i \leftarrow \rho_i
\rightarrow h_i$. Then
$$ F_1 F_2 = \eps(\rho_1,\rho_2) \cdot g_1\inv(F_2)h_2(F_1) \quad\hbox{if}\quad
O_2<O_1$$
and similar, exchanging $\eps_{12}$ with $\eps_{21}\inv$ and $h_i$ with 
$g_i$, if $O_1<O_2$.\footnote{Here and in the sequel we write $x < y$ (and
accordingly for sets) if the difference vector $x-y$ lies in the left 
space-like wedge $W_- = \{x: -x^1 > \vert x^0\vert\} = -W_+$.} \rm

The new feature in these commutation relations are the sector-dependent  
transformations under the broken symmetry, due to the non-triviality of the 
soliton automorphisms on space-like complements. They contribute extra phase 
factors $\pair{\chi_1}{h_2}/\pair{\chi_2}{g_1}$ to the exchange coefficient 
between two fields of given character.

Applying the commutation relations repeatedly, one obtains (with Lemma 2 (i))

{\bf Proposition 3:} (Weak Locality) \sl Let $F_i$ with soliton
charge $(g_i,h_i)$ and character $\chi_i$
be localized in $O_n < \ldots < O_1$. Then 
$$ F_1 \cdots F_n = {\prod}_{i<j} \eps_{ij} {\prod}_i \pair{\chi_i}{f_i} \cdot
F_n \cdots F_1 $$
where $\eps_{ij} = e(k_i,k_j)$, $k_i = h_ig_i\inv$, and 
$f_i = \prod_{n<i}g_n\inv
\prod_{i<n}h_n$. If $O_1 < \ldots < O_n$, then a similar law holds, 
exchanging $\eps_{ij}$ with $\eps_{ji}\inv$ and $h_i$ with $g_i$.

In particular, let $\Omega_\a$ be any of the vacuum vectors in
$H$, and $F_i$ with soliton charge $(g,h)$ and character $\chi_i$
be localized in $O_2 < O_1$. Then the two-point function satisfies
$$ (F_1\Omega_\a,F_2\Omega_\a) = \frac 1{\kappa_{\rho_{g,h}}}
\frac{\pair{\chi_2}g}{\pair{\chi_1}h} \cdot (F_2^*\Omega_\a,F_1^*\Omega_\a). $$
\rm

It is crucial to note here, that the character is {\sl not superselected} 
due to the spontaneous breakdown of the symmetry, that is,
vacuum correlations between fields of different character need not vanish.
If they did, then only operators of trivial character (neutral operators)
could have a vacuum expectation and consequently the vacuum states 
$\omega_\a$ were all equal. On the contrary, soliton operators of any given 
character will create dense subspaces of $H_{g,h}$, and these subspaces
are disjoint but not orthogonal.

\section{Spin-Statistics and CPT}

We want to prove Spin-Statistics and CPT theorems in terms of non-local 
Wightman fields associated with the soliton algebra. For this purpose,
we have to {\sl assume} that non-local Wightman fields $\varphi(x)$
are associated with the soliton algebra in the sense that there are 
limits of operators $F$ of point-like localization which are then translated 
to obtain $\varphi(x)$ (as unbounded operator-valued distributions, that is, 
the limits need to exist only after smearing with a test function). 
These fields of course inherit the Poincar\'e
and symmetry transformations as well as the * operation and the
commutation relations described by Prop.\ 2. Their quantum
numbers are the soliton charge $(g,h)$ and the character $\chi$, the 
latter not being superselected. The field $\varphi^*(x)$ carries the
charge $(g\inv,h\inv)$ and has character $\chi\inv$.

By appropriate projections, we may assume these fields to have 
definite spin $s \in \RR$, that is to transform as
$$ V(t) \varphi(x) V(t)\inv = e^{st} \varphi(\lambda x) $$
under the boost $\lambda = \pmatrix{\cosh t & \sinh t \cr \sinh t & 
\cosh t}$. This assumption requires implicitly the validity of 
condition (B) since the discussion of Lemma 1 shows that otherwise the
character would change continuously under $\Ad V(t)$. Clearly, $\varphi$ and 
$\varphi^*$ have the same spin.

To summarize, we shall assume:

(D) \sl There are sufficiently many covariant non-local Wightman fields 
$\varphi$ with quantum numbers $(g,h)$ and $\chi$ so that the vectors 
$\varphi(f)\Omega_\a$ span a dense subspace of $H_{g,h}$. \rm

The Reeh-Schlieder theorem applies to these fields, hence
also the fields localized in some region $O$ generate dense subspaces.

We cannot give convincing arguments on the validity of this assumption,
but they are certainly the best one may expect. A priori there is no reason
why soliton automorphisms should not be localizable in arbitrarily narrow
interpolation regions, but the existence of limits is certainly a nontrivial
dynamical problem. We refer to \cite{FJ} for the
analogous problem in scale invariant models.

{\bf Proposition 4:} (Spin-Statistics) \sl The spin $s \in \RR$ of a field
$\varphi$ in the sector $g\leftarrow\rho\rightarrow h$ and with character 
$\chi$ is
determined up to integers by
$$ e^{2\pi i s} = \kappa_\rho \cdot \pair\chi{hg\inv} $$ 
which depends only on $hg\inv$ and $\chi$. 

Proof: \rm
Standard arguments of Wightman theory, exploiting the spectrum condition
for the translations, tell that $V(t)$ can be analytically continued in
$t$ to positive resp.\ negative imaginary parts on (improper) vectors
$\varphi(x)\Omega_\a$ provided $x$ lies in the right resp.\ left space-like
wedge. For $\varphi_i$ both in the sector $(g,h)$ and $x < 0 < y$ it follows 
that
$$ \bea{l} (\varphi_1(x)\Omega_\a,\varphi_2(y)\Omega_\a) = \cr\quad = 
(V(-i\pi)\varphi_1(x)\Omega_\a,V(+i\pi)\varphi_2(y)\Omega_\a) = \cr\quad =
e^{i\pi(s_1+s_2)} (\varphi_1(-x)\Omega_\a,\varphi_2(-y)\Omega_\a) = \cr\quad =
\omega_{12} \cdot (\varphi_2(-y)^*\Omega_\a,\varphi_1(-x)^*\Omega_\a) . \eea $$
In the last step Weak Locality was used, and
$$\omega_{12} = e^{i\pi(s_1+s_2)} \frac 1{\kappa_\rho} \frac{\pair{\chi_2}g}
{\pair{\chi_1}h}. $$ 

The two-point function on the very right of this equation equals 
$(\varphi_2(x)^*\Omega_\a,\varphi_1(y)^*\Omega_\a)$ by translation invariance.
Especially for $\varphi_1 = \varphi_2$, both two-point functions on the very 
left and on the very right are positive distributions in $x-y$ satisfying
the spectrum condition. This is only possible if they are equal and the 
diagonal coefficient $\omega_{11}$ equals 1. \qed

Note that the spin is not determined by the sector alone \cite{SS},
and that fields of spin not differing by an integer may have
non-vanishing vacuum correlations since the character is not superselected. 

If we do the analogous computation leading to $\omega_{12}$ in the proof
of Prop.\ 4 with $y < 0 < x$ rather than $x < 0 < y$ we obtain another phase 
factor differing from $\omega_{12}$ 
by the substitutions $(s,\kappa,h) \leftrightarrow (-s,\kappa\inv,g)$. 
By the Spin-Statistics theorem, both expressions coincide. 
By translation invariance, the same relation between two-point functions 
$$\bea{l} (\varphi_1(x)\Omega_\a,\varphi_2(y)\Omega_\a) = \cr\qquad\qquad = 
\omega_{12} \cdot (\varphi_2(-y)^*\Omega_\a,\varphi_1(-x)^*\Omega_\a) \eea $$
therefore holds for all space-like $x - y$.

We now make a choice for the sign of $\mu(k):= \sqrt{e(k,k)}$ for each 
$k$, invariant under $k \leftrightarrow k\inv$, hence
$\mu(hg\inv)^2 = \kappa_{\rho_{g,h}}$. If $e(k_1,k_2)$ is a coboundary, 
we choose $\mu(k)$ as in Lemma 2 (ii). For every 
field $\varphi$ with quantum numbers $(g,h)$, $\chi$ and spin $s$ we define a 
complex phase 
$$\bea{r} \omega_\varphi := e^{i\pi s}\pair\chi{h\inv}/\mu(hg\inv) \cr
\equiv e^{-i\pi s}\pair\chi{g\inv}\mu(hg\inv) \eea $$ 
Clearly, $\omega_\varphi = \omega_{\varphi^*}$. 
Using the Spin-Statistics theorem, one easily verifies that 
$$\omega_{12} = \omega_{\varphi_1}/\omega_{\varphi_2}. $$ 

In view of the above relations between two-point functions, this means that 
all inner products between the (improper) vectors 
$\omega_\varphi \varphi(-x)^*\Omega_\a$ are the complex conjugates of those 
between $\varphi(x)\Omega_\a$. 

Hence we have the first part of

{\bf Proposition 5:} (CPT) \sl The antilinear operator $\Theta$ 
densely defined by 
$$ \Theta\varphi(x)\Omega_\a = \omega_\varphi \cdot \varphi(-x)^*\Omega_\a $$ 
is involutive and anti-unitary on its domain of definition and therefore
extends to an anti-unitary involution $\Theta$ on $H$. 

The CPT operator $\Theta$ transforms a
field $\varphi$ of soliton charge $(g,h)$ as follows:
$$ \Theta \varphi(x) \Theta\,\vert_{\textstyle H_{k,l}} = \zeta
 \cdot
(kl)\inv \left(\omega_\varphi \varphi(-x)^*\right) $$
with a prefactor
$\zeta = e(gh\inv,lk\inv)\inv
\frac{\mu(gh\inv lk\inv)}{\mu(gh\inv)\mu(lk\inv)}$.

If $e(k_1,k_2)$ is a coboundary (automatic for, e.g., $G=\ZZ$ or $\ZZ_N$) 
and hence $\zeta$ is trivial, then 
$$ \Theta B(W_\pm) \Theta = \Y B(W_\mp)\Y $$
where the unitary and involutive operator $\Y$ equals $Y_{(kl)\inv}$ on 
each subspace $H_{k,l}$. \rm 

The proof of the last statements will be given after Prop.\ 6. 

The rest of the discussion of CPT follows standard reasoning (e.g., 
\cite{BW,BSM,FRS2}). We call $\Mu$ and $\Kappa = \Mu^2$ the 
unitary diagonal operators with eigenvalues $\mu(hg\inv)$ and 
$\kappa_{\rho_{g,h}}$ on $H_{g,h}$, respectively, and compute (for $\varphi$
with soliton quantum numbers $(g,h)$)
$$ \bea{c} \Theta V(i\pi) \varphi(x)\Omega_\a = \Mu^*
h(\varphi(x)^*)\Omega_\a \quad (x>0) \cr
\Theta V(-i\pi) \varphi(x)\Omega_\a = 
\Mu \,g(\varphi(x)^*)\Omega_\a \quad (x<0). \eea $$
This suggests the introduction of antilinear densely defined unbounded 
operators by
$$ \bea{c} \dot S_+ F \Omega_\a := h(F^*)\Omega_\a \qquad (F\in B(W_+)) \cr
\dot S_- F \Omega_\a := g(F^*)\Omega_\a \qquad (F\in B(W_-)) \eea $$
if $F$ carries soliton charge $(g,h)$.

By Weak Locality, one sees that $\dot S_\pm^*$ are defined on the domains of 
$\dot S_\mp$ where they coincide with the latter up to a unitary factor:
$\dot S_\pm^* \supset \Kappa^{\pm1} \dot S_\mp$. Thus they are closable 
and the closures $S_\pm$ possess unique polar decompositions. 
By comparison, and since $V(\pm i\pi)$ are positive operators, one gets

{\bf Proposition 6:} (Modular Structure) \sl The closures of $\dot S_\pm$ 
satisfy 
$$ S_\pm^* = \Kappa^{\pm1} \cdot S_\mp$$  
and have polar decompositions
$$ S_\pm = \Mu^{\pm 1}\Theta \cdot V(\pm i\pi). $$ 
The following commutation relations with boosts and translations hold:
$$ \bea{c} \Theta V(\pm i\pi) = V(\mp i\pi) \Theta \cr
\Theta V(t) = V(t) \Theta \cr
\Theta U(x) = U(-x) \Theta. \eea $$

Proof of the commutation relations: \rm The first relation obtains from
$S_+ = \Kappa S_-^*$. The second one follows since $V(t)$ is an imaginary 
power of $V(i\pi)$. The last one follows from the definition of $\Theta$. \qed

\sl Proof of the CPT transformation law (Prop.\ 5): \rm We compute the 
improper vector $\Theta \varphi(x)\varphi_1(y)\Omega_\a$ where 
$\varphi_1$ belongs to the sector $(k\inv,l\inv)$. E.g., for $0 < x < y$ 
where $V(i\pi)$ is defined, we substitute $\Theta = \Mu S_-V(i\pi)$ 
and apply the definitions as well as the commutation relations of Prop.\ 2 to 
restore the operator order inverted by $S_-$. The result is proportional to
$\varphi(-x)^*\Theta\varphi_1(y)\Omega_\a$. The accumulating phase factors
are seen, using the Spin-Statistics theorem, to give rise to $\zeta$
as in the proposition. The twist by $\Y$ removes the symmetry
transformation from the transformation law, and (if $\zeta = 1$)
$$ \Y\Theta \varphi(x) \Theta \Y = \omega_\varphi \cdot 
Y_{gh}\varphi(-x)^*. $$
The image is a field with soliton charge $(h,g)$ if $\varphi$ has charge
$(g,h)$. \qed

We note, however, that one crucial point of the Bisognano-Wichmann result 
\cite{BW} for local Wightman fields is missing. Namely, our operators $S_\pm$ 
are {\sl not} Tomita operators $F\Omega \mapsto F^*\Omega$ ($F \in B(W_\pm)$)
defined by the 
adjoint alone, but involve the symmetry transformations. Since the latter do 
not preserve the reference vacuum vector, $S_\pm$ are also not related by 
unitaries to a Tomita operator, and therefore its polar factor $\Theta$ 
cannot be unitarily related with the Tomita conjugation $J$. As $J$ takes the
respective algebras into their commutants, the image of $\Theta$ is {\sl not} 
a twisted commutant. In view of the CPT transformation law in Prop.\ 5, 
even if there is no cohomological obstruction, we conclude that the soliton 
algebras of the two wedges are {\sl not} their mutual twisted commutants with 
any unitary twist operator.

The present CPT operator $\Theta$ takes a given sector $(g,h)$ into its 
conjugate sector $(g\inv,h\inv)$. This is not precisely what one expects
from a CPT operator which should rather exchange the two asymptotic
directions and therefore connect $(g,h)$ with $(h,g)$. Furthermore, as one
might have already remarked, $S_\pm$ depend on the choice of the reference 
vacuum vector $\Omega_\a$. E.g., on $H_\a$ they coincide with the Tomita
operators $\pi_\a(a)\Omega_\a \mapsto \pi_\a(a^*)\Omega_\a$ ($a \in A(W_\pm)$)
for this vacuum sector, but other vacuum sectors $H_{\a\comp k}$
are mapped into $H_{\a\comp k\inv}$. 

Both these unpleasant features can be amended by a trivial trick: One changes
$S_\pm$ and $\Theta$ into $\Y S_\pm=S_\pm \Y$ and $\Y\Theta=\Theta \Y$ 
where $\Y\vert_{\textstyle H_{g,h}} = Y_{(gh)\inv}$. Since 
$\Y$ is a unitary 
involution on $H$ and commutes with $\Mu$, the polar decomposition 
formulae are not affected by this change. The new operators map $H_{g,h}$ 
into $Y_{gh}H_{g\inv,h\inv} = H_{h,g}$ as desired, and one can convince 
oneself that the action of the $\Y S_\pm$ on the entire Hilbert 
space, albeit defined relative to some reference vacuum vector, no longer
depends on the choice of the latter. In particular, they 
coincide with the Tomita operators resp.\ the CPT operator of the
observables on each of the vacuum sectors simultaneously. 

As a bonus, the modified CPT operator $\Y\Theta$ maps the opposite wedge 
algebras directly into each other, without a twist (cf.\ Prop.\ 5).

\section{Neutral solitons}

If it were not for the option (cf.\ the discussion at the end of the previous 
section) to have $(h,g)$ along with $(g,h)$ among the sectors, one can as 
well define the ``right-oriented''
soliton algebra $B_R \subset B$ generated by the operators with soliton charge 
$\rho_{e,h}$ only. It acts irreducibly on each of the 
subspaces $H_{k,R} = \bigoplus_h H_{k,h}$ which 
contain only one vacuum sector $H_\b$, $\omega_\b = \omega_\a\comp k$, each. 
On $B_R$ the dual symmetry reduces to $\hat G$. The $G$-symmetry (no longer 
inner) is again broken and connects the representations on $H_{k,R}$. 

The entire discussion leading to Props.\ 3--6 remains valid for $B_R$ 
on every single $H_{k,R}$, putting all left vacuum quantum numbers $g=e$.
Only the operator $\Y$ in Prop.\ 5 cannot be defined on $H_{k,R}$.

The right-oriented soliton algebra is completely satisfactory if one is
interested in the superselection sectors of the {\sl neutral} observables
$A^0$. By condition (A), the automorphisms $\rho_{g,h}$ restrict on $A^0$ 
and turn into DHR automorphisms since the asymptotic transformations under 
$G$ become trivial.
Furthermore, the restrictions of $\rho_{g,h}$ and $\rho_{gk,hk}$ coincide.
Therefore, the restricted automorphisms $\rho_{e,h}$ exhaust the DHR
sectors of $A^0$. 

We emphasize that the vacuum sectors of $A$ not only become equivalent upon
restriction to $A^0$, but also remain irreducible. This fact is due to
the spontaneous breakdown, and implies that $A^0$ is not Haag-dual  
in its vacuum representation. This explains why $A^0$ can possess DHR sectors, 
in view of M\"uger's result \cite{MM2} on the absence of DHR sectors
in Haag-dual massive (that is: SPW) theories.

The field algebra $B^0$ for $A^0$ is generated by operators 
$$ F = (\rho_{e,h},b) \qquad (b \in A^0) $$
which are neutral with respect to $G$ but carry a DHR charge,
with the same localization and * structures as described for $B$ in Sect.\ 3. 
It is clear that this field algebra equals the neutral subalgebra 
$$ B^0 = (B_R)^G $$
of the right-oriented soliton algebra. Note that the subalgebra $B^0$ is 
preserved by the Poincar\'e group only if the cocycles for the boosts are 
neutral (condition (B)).

One can again repeat the entire discussion of Sect.\ 3 and derive 
Props.\ 3--6 for $B^0$. The difference is now that all fields involved are 
neutral, that is, all characters and transformations under $G$ are trivial. 
Thus, the spin is indeed determined by the sector alone:
$$ e^{2\pi i s} = \kappa_\rho$$
according to Prop.\ 4, and is superselected mod $\ZZ$.

Furthermore, the restriction of $\Theta$ is the true CPT operator
(provided $\zeta = 1$ in Prop.\ 5)
$$ \Theta B^0(W_\pm) \Theta = B^0(W_\mp), $$ 
and the restricted operators $S_\pm$ {\sl are} the 
Tomita operators $F\Omega \mapsto F^*\Omega$ ($F \in B^0(W_\pm)$). Hence
$$ \Mu^{\pm1}\Theta = J_\pm $$
are the Tomita conjugations of the right and left wedge algebras $B^0(W_\pm)$
with respect to the vacuum vector, which take the respective 
algebras into their commutants.

We conclude for the neutral soliton algebra:

{\bf Proposition 7:} (Twisted Duality) \sl If $e(k_1,k_2)$ is a coboundary, 
then the algebra $B^0(W_-)$ of the left wedge equals 
the twisted commutant of the algebra $B^0(W_+)$ of the right wedge, that is:
$$ B^0(W_-) = \Theta B^0(W_+) \Theta = 
\Mu\inv B^0(W_+)' \Mu $$ 
where the twist operator $\Mu$ is a square root of $e^{2\pi i s}$. \rm

\section{Discussion}
Let us consider the square of inclusions 
$$ \matrix{B_R & \supset & B^0 \cr \cup && \cup \cr A & \supset & A^0} $$
where the horizontal inclusions are fixpoints under the broken $G$-symmetry
and the vertical inclusions are fixpoints under the dual $\hat G$-symmetry.

The theories in the same row live on the same Hilbert space, while the
vertical extensions require an extension of the Hilbert space.

The net $A^0$ violates Haag duality, while $A$ was assumed to be Haag dual. 
Therefore, $A$ is the dual net canonically associated with $A^0$. The net 
$B^0$ was shown to satisfy twisted duality (Prop.\ 7) while nothing of the 
sort could be shown for $B_R$. The reason is clear: $B^0$ being already 
twisted dual, it cannot possess any twisted dual extension on the same 
Hilbert space.

A comparison suggests itself with the situation studied by M\"uger 
\cite{MM1}
$$ \matrix{\hat F & \supset & F \cr \cup && \cup \cr A^d & \supset & A} $$
where $F$ was assumed to satisfy (fermionic) duality with an unbroken
symmetry $G$ with fixpoints $A$. The fixpoints then violate duality, and 
M\"uger constructed the dual net $A^d$ by first extending $F$
to a non-local net $\hat F$ including ``disorder operators'' which implement 
the symmetry of $F$ on the left complement of some region and leave invariant 
the elements of $F$ in the right complement. (For the existence of such
operators, the assumption of the split property for wedges (SPW) is 
essential). He showed that $\hat F$ carries a dual symmetry under $\hat G$ 
which is spontaneously broken and commutes with the original symmetry 
(actually, in the non-abelian case, $G$ and $\hat G$ together form the 
quantum double). The dual net $A^d$ is then obtained as the fixpoints of 
$\hat F$ under the original symmetry.

The two scenarios under discussion obviously describe the same situation,
by identifying the diagrams and exchanging the roles of $G$ and its dual. 
The difference consists in the circumstance that we construct the
``top right'' net from the ``bottom left'' net, passing via the ``top left'' 
non-local net, while the analysis in \cite{MM1} goes the opposite way.
The latter point of view is more
general as it admits non-abelian unbroken groups, while ours
is more general as $B^0$ needs not to be fermionic but rather may turn
out to be anyonic. 

To complete the comparison, we should be able to describe the
disorder operators \cite{MM1} in our setting. Indeed, they are given by
$U_\eta = (id,u) \in A$ where $u$ is a local unitary of character $\eta$. By 
Prop.\ 2, $U_\eta F U_\eta^*$ equals $F$ if $F \in B^0$ is localized 
at right space-like distance from $U_\eta$, and equals $\eta(F)$ if $F$ is 
localized at left space-like distance from $U_\eta$. It is clear that the 
unitaries $U_\eta$ along with the neutral operators in $B^0$ generate $B_R$.

As an instructive example which is perfectly consistent with the picture,
we want to match it with our semi-classical and perturbative knowledge of the 
Sine-Gordon vs.\ massive Thirring model \cite{SC}. Let
$A$ denote the Sine-Gordon model generated by its scalar field $\phi$
which carries the broken $\ZZ$-symmetry $\phi \mapsto \phi+2\pi n$.
The fixpoints $A^0$ are the fields $\partial\phi$ and $\sin\phi$, $\cos\phi$.
The solitons of the Sine-Gordon model are known to be the Thirring fermions 
$\psi$, therefore the extension $B_R$ by the solitons should be a theory 
containing both fields $\phi$ and $\psi$ with complicated commutation 
relations and both the $\ZZ$-symmetry of the Sine-Gordon model and the 
$U(1)$-symmetry of the Thirring model. Passing to the fixpoints under $\ZZ$ 
eliminates the field $\phi$ but retains the derivative and trigonometric 
fields which are identified \cite{SC} with the quadratic gauge-invariant 
combinations of the Thirring fermion. Therefore, $B^0$ is just the Thirring 
model. Its fixpoints under $U(1)$ are generated by the quadratic invariants, 
hence coincide with $A^0$. This completes the square of inclusions.

Note that this example is both abelian, and yields a fermionic theory $B^0$.
It thereby belongs to the intersection of models covered by M\"uger's 
and by our approach, respectively.

The departure from the standard form of the fundamental theorems will indeed 
be seen in this familiar model as soon as one considers ``mixed''
fields \cite{SS} in
the style of $\colon \psi \exp i\a\phi \colon$, which involve both Thirring 
fermions and Sine-Gordon fields of non-trivial character.

\vskip5mm

{\bf Acknowledgments:} This work has been done at the 
Werner-Heisenberg-Institut (MPI), M\"unchen, and at the 
Erwin Schr\"odinger International Institute (ESI), Vienna. To both 
institutions I am indebted for hospitality and for financial support. 
Many thanks go also to M. Niedermaier, M. M\"uger, D. Schlingemann and 
others for stimulating discussions. 

%\vfill

\small \baselineskip=10.5pt \parskip=1.0mm minus0.3mm

\end{document}